\documentclass[twocolumn]{aa}
\usepackage{rotating}
\usepackage{graphicx}

\font\twlbfs=cmbxsl10 scaled \magstep2

\def\ea{et al. }

\begin{document}
\headnote{Research Note}

\title{New Galactic Wolf-Rayet stars, and candidates}

\subtitle{An annex to {\twlbfs The VIIth Catalogue of Galactic Wolf-Rayet Stars}}

\author{Karel A. van der Hucht
          \inst{1,2}
       }

\offprints{$<$K.A.van.der.Hucht@SRON.nl$>$}

\institute{SRON Netherlands Institute for Space Research,
           Sorbonnelaan 2, NL-3584CA Utrecht, the Netherlands \\
           \email{k.a.van.der.hucht@sron.nl}
   \and
           Astronomical Institute Anton Pannekoek,
           University of Amsterdam,
           Kruislaan 403, 1097 SJ Amsterdam, the Netherlands  \\
          }

\date{Received 13 June 2006 / Accepted 13 July 2006}

\abstract{
This paper gathers, from the literature and private communication, 72 new
Galactic Population~I Wolf-Rayet stars and 17 candidate WCLd stars,
recognized and/or discovered after the publication of {\it The VIIth
Catalogue of Galactic Wolf-Rayet Stars}. This brings the total number of 
known Galactic Wolf-Rayet stars to 298, of which 24 (8\,\%) are in open 
cluster Westerlund\,1, and 60 (20\,\%) are in open clusters near the 
Galactic Center.
 }

\maketitle
%
%________________________________________________________________

\section{Introduction}

Wolf-Rayet (WR) stars represent the final phase in the evolution of
massive stars (i.e., $M_{\rm i}$\,$\ga$\,20\,M$_\odot$), before becoming
a supernova and/or stellar remnant.  They are the chemically evolved 
descendants of OB stars (e.g., Meynet \& Maeder \cite{mey2005}) and 
contribute to chemical and kinetic enrichment of their environment through 
their dense stellar winds and Lyman continuum photons. Some of them could 
be the possible progenitors of core-collapse supernovae and $\gamma$-ray 
bursts, especially in a low metallicity environment (e.g., Hirschi \ea 
\cite{hir2005}; Petrovic \ea \cite{pet2005}; Yoon \& Langer \cite{yoo2005};
Langer \& Norman \cite{lan2006}; Woosley \& Heger \cite{woo2006};
Fruchter \ea \cite{fru2006}).  Where $\sim$\,35\% of the
Galactic WR stars have wind blown bubbles, visible as ring nebulae (Marston
\cite{mar1997}), they provide the ideal environment for a $\gamma$-ray
burst afterglow (e.g., Chevalier \cite{che2005}; Dwardakas \cite{dwa2005};
Zou \ea \cite{zou2005}; Eldridge \ea \cite{eld2006a}; Eldridge \& Vink
\cite{eld2006b}; Hammer \ea \cite{ham2006}).  For all practical purposes, it
is important to know as many WR stars as possible. Assembling a complete 
catalogue of WR stars, their spectral types (and hence chemical make-up) 
and relative numbers is important in order to understand their impact on 
the Galactic environment as well as to investigate their suitability as 
precursors to very energetic processes in extragalactic systems.

Since the publication of {\it The VIIth Catalogue of Galactic Wolf-Rayet
Stars} (van der Hucht \cite{huc2001}, henceforth 7Cat), numerous new
Galactic Population\,I Wolf-Rayet (WR) stars have been discovered, notably 
near the Galactic Center (in the infrared) and in open clusters (e.g., in 
Westerlund\,1, optically), but also as individual field stars, thanks 
to the advancements in sensitivity and spatial resolution.  In order to list
these new WR stars properly in the 7Cat numbering system, and because of
the crowding and the occasional resolution of apparently single objects
into multiple objects, it became necessary to have the RA/Dec(J2000) 
coordinates of the 26 7Cat WR stars near the Galactic Center re-determined 
with higher accuracy.  For example, with improving 
spatial resolution and sensitivity, it appears that what Krabbe \ea
(\cite{kra1995}) saw as the single object GC\,IRS\,13E (= WR\,101f in 7Cat,
WN9-10), has been resolved by Maillard \ea (\cite{mai2004}) into a cluster
containing 7 stars, including two WR stars (GC\,IRS13\,E2 and
GC\,IRS13\,E4) and three candidate WCLd stars (GC\,IRS13\,E3A,
GC\,IRS13\,E3B, and GC\,IRS13\,E5).  A critical analysis of IRS\,13E has
been presented recently by Paumard \ea (\cite{pau2006}).

This paper, rather than providing a completely revised WR catalogue,
presents as an Annex to the 7Cat a list of new WR stars and candidate WR
stars discovered in recent years, together with updated coordinates for
some objects.

All of the new discoveries quoted here require confirmation by additional
multi-frequency high-spectral resolution and high-angular resolution 
observations, which may throw new light on earlier results, e.g., Tanner 
\ea (\cite{tan2006}). For example, GC\,IRS8, one of five GC stars
suggested by Tanner et al. (\cite{tan2005}) to be WR stars, turned out
to be an O5-O6 giant or supergiant when observed by Geballe \ea
(\cite{geb2006}).

%__________________________________________________________________

\section{New data}

The new Galactic WR stars listed in this Annex have been discovered by
the following authors:

\smallskip \noindent
- Some 15 possible WR stars in the Arches cluster had been recognized
(in the infrared) before 2001 by Nagata \ea (\cite{nag1995}) and Cotera 
\ea (\cite{cot1996}), as noted by Blum \ea (\cite{blu2001}), Lang \ea 
(\cite{lan2001}), and Figer \ea (\cite{fig2002}).  Those objects are 
now included in this Annex.

\smallskip \noindent
- Bartaya \ea (\cite{bar1994}) discovered (in the optical) one new WN4 star 
(WR\,159) in the OB association Cas\,OB\,4, which has been re-discovered by 
Negueruela (\cite{neg2003}).

\smallskip \noindent
- Figer \ea (\cite{fig1996}) discovered (in the infrared) two WN9/Ofpe and 
five possible WCLd stars in the Quintuplet cluster.  Tuthill \ea (\cite{tut2006}) 
showed that at least two of those WCLd stars, Q2 and Q3, have infrared pinwheels, 
indicative of dust formation originating in the colliding winds of long period 
WCL+OB binaries.

\smallskip \noindent
- Clark \& Negueruela (\cite{cla2002}), Negueruela \& Clark (\cite{ncl2003}), 
Clark \ea (\cite{cla2005}), Negueruela \& Clark (\cite{ncl2005}), Negueruela 
(priv.  comm.), and Crowther \ea (\cite{cro2006}), discovered (in the optical) 
24 new WR stars in the open cluster Westerlund\,1, extremely rich in O stars, 
WR stars and LBVs.  Groh \ea (\cite{gro2006}) independently discovered (in the 
optical) three W\,d1 WR stars discovered also by Crowther \ea (\cite{cro2006}). 

\smallskip \noindent
- Pasquali \ea (\cite{pas2002}) discovered (in the infrared) one new WC8 
star (WR\,142a), in Cygnus.

\smallskip \noindent
- Homeier \ea (\cite{hom2003}) discovered (in the infrared) three WC8-9 stars 
and one WN10 star, in the inner Galaxy.

\smallskip \noindent
- Drew \ea (\cite{dre2004}) discovered (in the optical) one new WO3 star 
(WR\,93b), located most likely in the Scutum-Crux arm of the inner Milky Way, 
from follow-up observations of candidate emission-line stars in the AAO/UKST 
Southern Galactic Plane H$\alpha$ Survey (Parker \ea \cite{par2005}).

\smallskip \noindent
- Cohen \ea (\cite{coh2005}) discovered (in the optical and infrared) one new 
WN7 star (WR\,75ab), from follow-up observations of candidate emission-line 
stars in the AAO/UKST Southern Galactic Plane H$\alpha$ Survey (Parker \ea 
\cite{par2005}).

\smallskip \noindent
- Hopewell \ea (\cite{hop2005}) discovered (in the optical) five new WC9 stars 
in a programme of follow-up optical spectroscopy of candidate emission-line
stars in the AAO/UKST Southern Galactic Plane H$\alpha$ Survey (Parker \ea
\cite{par2005}).

\smallskip \noindent
- Paumard \ea (\cite{pau2001}, \cite{pau2006}), Eckart \ea
(\cite{eck2004}), Horrobin \ea (\cite{hor2004}), Maillard \ea
(\cite{mai2004}), Moultaka \ea (\cite{mou2005}), and Tanner \ea
(\cite{tan2002}, \cite{tan2005}) together discovered (in the infrared)
14 new WR stars and 14 candidate WCLd stars in the Galactic Center cluster.

\smallskip \noindent
- Eikenberry \ea (\cite{eik2001}, \cite{eik2004}) discovered (in the infrared)
one new WC9 star (WR\,111b) in the cluster apparently near the soft 
$\gamma$-ray repeater SGR\,1806$-$20.

\smallskip \noindent
- Figer \ea (\cite{fig2005}) discovered (in the optical and infrared)
three new WR stars in the cluster
around the soft $\gamma$-ray repeater SGR\,1806$-$20, two of which
(WR\,111a and WR\,111c) had been discovered independently (in the infrared)
by J. LaVine and S.S. Eikenberry (2004, private communication).

%__________________________________________________________________

\section{The census of Population\,I WR stars and candidate
         WR stars in the Galaxy}

The new Galactic WR stars and candidates are listed in Table\,1, together
with those WR stars from the 7Cat for which the coordinates have been
re-determined. Table\,1 lists:                    \\
- Galactic WR running number in the 7Cat system;  \\
- WR discovery designation, acknowledging the authors  \\
  \hspace*{2mm} of the discovery paper;              \\
- additional designation(s) from the literature;  \\
- discovery spectral type;                        \\
- revised spectral type;                          \\
- magnitude ($V$, or $R$, or $K$);                \\
- RA/Dec(J2000) coordinates;                      \\
- discovery reference.                            

\bigskip

% \noindent
% ============ \\
% TABLE 1      \\
% ============ \\

There are 72 new Galactic WR stars listed in this Annex, plus 17
candidate WR stars, some possibly of the WCLd type.  Of the 72 new
Galactic WR stars we find: 45 WN stars, 26 WC stars, and one WO star.

Of the 72 new Galactic WR stars, in most cases the number of observations
is still too small to establish which are binaries. We would
expect a binary frequency of $\sim$\,40\% (van der Hucht, \cite{huc2001},
Table\,20).  Only a few new WR stars have shown some indication of binarity 
(see Table\,1).

There are now 60 known WR stars in the open clusters near the Galactic
Center, i.e., the Galactic Center cluster (29, plus 13 candidate WR stars),
the Arches Cluster (17, all WN) and the Quintuplet cluster (14, plus 3
candidate WR stars), plus 16 candidate WR stars, mostly candidate WCLd.  

\bigskip

Together with the 226 WR stars in the 7Cat, this Annex brings the total
number of presently known Galactic Population\,I WR stars to 298, excluding
the 17 candidate WR stars.  The spectral subtype distribution is: 
171 WN stars, 10 WN/WC stars, 113 WC stars, and 4 WO stars.

The 7Cat has 53 of its 226 WR stars in open clusters and OB associations,
i.e. 23\,\%.  Together with this Annex we count 137 out of the 298 known
Galactic WR stars in open clusters and OB associations, i.e. 46\,\%, of
which 8\,\% are in open cluster Westerlund\,1 and 20\,\% are in open 
clusters near the Galactic Center.    

\vspace*{-3cm}

%__________________________________________________ 12 columns table, sideways

\begin{center}
\begin{table*}[t!]
\begin{sideways}
\begin{tabular}{| l   l                l                                      | l                       l             l    | r              l     | l                     l                       c     | l           |}
\hline\hline
\multicolumn{12}{|l|}{}                                                                                                                                                                                                       \\[-1.5mm]
\multicolumn{12}{|l|}{{\bf Table 1:} {\large New Galactic Wolf-Rayet stars.
 {\it Data quoted from 7Cat are listed in italics},
 revised and new data are listed in roman font.}}                                                                                                                                                                                     \\
\multicolumn{12}{|l|}{}                                                                                                                                                                                                       \\[-1.5mm]
\hline\hline
                    &                &                                        &                       &             &      &              &       &                     &                       &       &             \\[-2.5mm]
  WR                &   WR           & other designation(s)                   &      discovery        &    revised  &      &$m$~~~~~~~~~~ &       & RA(J2000)           & Dec(J2000)            &       & WR          \\
                    &   discovery    &                                        &      spectral         &    spectral &      &(mag)~~~~~    &       &                     &                       &       & discovery   \\
                    &   designation  &                                        &      type             &    type     & ref. &              & ref.  &                     &                       & ref.  & ref.        \\
                    &                &                                        &                       &             &      &              &       &                     &                       &       &             \\[-2.5mm]
\hline\hline
                    &                &                                        &                       &             &      &              &       &                     &                       &       &             \\
     ~75aa          &      HBD\,1    &      SHS\,J162620.2$-$455946           &      WC9d             &             &      & $I$ = 14.18  & HB05  &      16 26 20.2     &      $-$45 59 46      & HB05  & HB05        \\[-3mm]
                    &                &                                        &                       &             &      &              &       &                     &                       &       &             \\
     ~75ab          &      CPG\,1    &                                        &      WN7h             &             &      & $K_s$= ~8.91 & CP05  &      16 33 48.74    &      $-$49 28 43.5    & CP05  & CP05        \\[-3mm]
                    &                &                                        &                       &             &      &              &       &                     &                       &       &             \\
     ~75c           &      HBD\,2    &      SHS\,J163403.6$-$434025           &      WC9              &             &      & $I$ = 13.12  & HB05  &      16 34 03.6     &      $-$43 40 25      & HB05  & HB05        \\[-3mm]
                    &                &                                        &                       &             &      &              &       &                     &                       &       &             \\
     ~75d           &      HBD\,3    &      SHS\,J163417.5$-$460852           &      WC9              &             &      & $I$ = 12.30  & HB05  &      16 34 17.5     &      $-$46 08 52      & HB05  & HB05        \\[-3mm]
                    &                &                                        &                       &             &      &              &       &                     &                       &       &             \\
\hline
                    &                &                                        &                       &             &      &              &       &                     &                       &       &             \\
\multicolumn{3}{|l|}{\large\it open cluster Westerlund\,1}                    &                       &             &      &              &       &                     &                       &       &             \\[+2mm]
     ~77aa          & HBD\,4, NC-T   &      SHS\,J164646.3$-$454758           &      WC9d             &             &      & $J$ = 10.04  & 2MASS &      16 46 46.3     &      $-$45 47 58      & HB05  & HB05        \\
     ~77a           &      NC-Q      &                                        &      WN6-7            &    WN6      & CH06 & $J$ = 11.72  & CH06  &      16 46 55.55    &      $-$45 51 35.0    & CH06  & NC05        \\
     ~77b           &      NC-N      &                                        &      WC8              &    WC9d     & CH06 & $J$ = ~9.69  & 2MASS &      16 46 59.9     &      $-$45 55 26      & 2MASS & NC05        \\
     ~77c $^1$      &      NC-I      &                                        &      WN6-8            &    WN8      & CH06 & $J$ = 10.89  & CH06  &      16 47 00.88    &      $-$45 51 20.8    & CH06  & CN02, NC03  \\
     ~77d           &      NC-P      &      Wd1-57c                           &      WN8              &    WN7      & CH06 & $J$ = 11.06  & CH06  &      16 47 01.59    &      $-$45 51 45.5    & CH06  & NC05        \\
     ~77e $^2$      &      NC-J      &                                        &      WNL              &    WN5      & CH06 & $J$ = 11.7:\,& CH06  &      16 47 02.47    &      $-$45 51 00.1    & CH06  & CN02, NC03  \\
     ~77f           &      NC-S      &      Wd1-5                             &      WNVL             &    WN10-11: & CH06 & $J$ = ~9.81  & CH06  &      16 47 02.98    &      $-$45 50 20.0    & CH06  & NC05        \\
     ~77g $^3$      &      NC-K      &                                        &      WC               &    WC8      & CH06 & $J$ = 11.81  & CH06  &      16 47 03.25    &      $-$45 50 43.8    & CH06  & NC03        \\
     ~77h           &      NC-V      &                                        &      WN8              &             &      & $J$ = 10.47  & CH06  &      16 47 03.81    &      $-$45 50 38.8    & CH06  & Ne05        \\
     ~77i           &      NC-M      &      Wd1-66                            &      WC9              &    WC9d     & CH06 & $J$ = 10.13  & CH06  &      16 47 03.96    &      $-$45 51 37.8    & CH06  & NC05        \\
     ~77j $^{\,4}$  &      NC-G      &                                        &      WN6-8            &    WN7      & CH06 & $J$ = 11.35  & CH06  &      16 47 04.01    &      $-$45 51 25.2    & CH06  & CN02, NC03  \\
     ~77k           &      NC-L      &      Wd1-44                            &      WN9              &    WN9:     & CH06 & $J$ = ~9.08  & CH06  &      16 47 04.19    &      $-$45 51 07.4    & CH06  & NC05        \\
     ~77l $^5$      &      NC-H      &                                        &      WC9              &    WC9d     & CH06 & $J$ = 10.3:\,& CH06  &      16 47 04.22    &      $-$45 51 20.2    & CH06  & CN02, NC03  \\
     ~77m $^{6}$    &      NC-C      &                                        &      WC8              &    WC9d     & NC05 & $J$ = 11.26  & CH06  &      16 47 04.40    &      $-$45 51 03.8    & CH06  & CN02, NC03  \\
     ~77n $^{7}$    &      NC-F      &      Wd1-239                           &      WC9              &    WC9d     & CH06 & $J$ = ~9.85  & CH06  &      16 47 05.22    &      $-$45 52 25.0    & CH06  & CN02, NC03  \\
     ~77o $^{~8}$   &      NC-B      &                                        &      WNL              &    WN7      & CH06 & $J$ = 10.91  & CH06  &      16 47 05.36    &      $-$45 51 05.0    & CH06  & CN02, NC03  \\
     ~77p $^{~9}$   &      NC-E      &      Wd1-241                           &      WC9              &             &      & $J$ = 10.12  & CH06  &      16 47 06.05    &      $-$45 52 08.2    & CH06  & CN02, NC03  \\
     ~77q           &      NC-R      &      Wd1-14c                           &      WN6-7            &    WN5      & CH06 & $J$ = 11.92  & CH06  &      16 47 06.07    &      $-$45 50 22.6    & CH06  & NC05        \\
     ~77r $^{10}$   &      NC-D      &                                        &      WN6-8            &    WN7      & CH06 & $J$ = 11.63  & CH06  &      16 47 06.24    &      $-$45 51 26.5    & CH06  & CN02, NC03  \\
                    &      NC-U      &                                        &      WN4              &    WN6      & CH06 & $J$ = 10.76  & CH06  &      16 47 06.55    &      $-$45 50 39.0    & CH06  & Ne05        \\[-2.4mm]
     ~77s~~~\, \{   &                &                                        &                       &             &      &              &       &                     &                       &       &             \\[-2.4mm]
                    &      GDTB\,1   &                                        &      WN5-7            &             &      & $K$ = ~9.19  & GD06  &      16 47 06.6     &      $-$45 50 38.6    & GD06  & GD06        \\
                    &      NC-W      &                                        &      WN5-6:           &    WN6      & CH06 & $J$ = 12.11  & CH06  &      16 47 07.58    &      $-$45 49 22.2    & CH06  & CH06        \\[-2.4mm]
     ~77sa~~   \{   &                &                                        &                       &             &      &              &       &                     &                       &       &             \\[-2.4mm]
                    &      GDTB\,3   &                                        &      WN7              &             &      & $K$ = ~9.70  & GD06  &      16 47 07.6     &      $-$45 49 21.7    & GD06  & GD06        \\
     ~77sb          &      NC-O      &                                        &      WN6              &             &      & $J$ = 11.00  & CH06  &      16 47 07.66    &      $-$45 52 35.9    & CH06  & NC05        \\
     ~77sc $^{\,11}$&      NC-A      &      Wd1-72                            &      WN4-5            &    WN7      & CH06 & $J$ = 10.34  & CH06  &      16 47 08.32    &      $-$45 50 45.5    & CH06  & CN02, NC03  \\
                    &      NC-X      &                                        &      WN4-5:           &    WN5      & CH06 & $J$ = 12.36  & 2MASS &      16 47 14.1     &      $-$45 48 32      & 2MASS & CH06        \\[-2.4mm]
     ~77sd~~   \{   &                &                                        &                       &             &      &              &       &                     &                       &       &             \\[-2.4mm]
                    &      GDTB\,2   &                                        &      WN4-5            &             &      & $K$ = ~9.99  & GD06  &      16 47 14.2     &      $-$45 48 31.4    & GD06  & GD06        \\
                    &                &                                        &                       &             &      &              &       &                     &                       &       &             \\
\hline
                    &                &                                        &                       &             &      &              &       &                     &                       &       &             \\
     ~77t           &      HBD\,5    &      SHS\,J165057.6$-$434028           &      WC9d             &             &      & $I$ = 13.00  & HB05  &      16 50 57.6     &      $-$43 40 28      & HB05  & HB05        \\[-3mm]
                    &                &                                        &                       &             &      &              &       &                     &                       &       &             \\
     ~93b           &      DBU\,1    &                                        &      WO3              &             &      & $K$ = 10.17  & DB04  &      17 32 03.30    &      $-$35 04 32.5    & DB04  & DB04        \\[-3mm]
                    &                &                                        &                       &             &      &              &       &                     &                       &       &             \\
\hline
\end{tabular}
\end{sideways}
\end{table*}
\end{center}

% \clearpage

\begin{center}
\begin{table*}[t!]
\begin{sideways}
\setlength{\tabcolsep}{1.7mm}
\begin{tabular}{|lll|lll|ll|llc|l|}
\hline\hline
\multicolumn{12}{|l|}{}                                                                                                                                                                                               \\[-1.5mm]
\multicolumn{12}{|l|}{{\bf Table 1 (cont'd):} {\large New Galactic Wolf-Rayet stars.
 {\it Data quoted from 7Cat are listed in italics},
 revised and new data are listed in roman font.}}                                                                                                                                                                                     \\
\multicolumn{12}{|l|}{}                                                                                                                                                                                               \\[-1.5mm]
\hline\hline
                    &                &                                             &                  &             &      &              &       &                     &                       &       &             \\[-2.5mm]
  WR                &   WR           & other designation(s)                        &      discovery   &    revised  &      &   $m$        &       & RA(J2000)           & Dec(J2000)            &       & WR          \\
                    &   discovery    &                                \hfill{in}   &      spectral    &    spectral &      &  (mag)       &       &                     &                       &       & discovery   \\
                    &   designation  &                                \hfill{PG06} &      type        &    type     & ref. &              & ref.  &                     &                       & ref.  & ref.        \\
                    &                &                                             &                  &             &      &              &       &                     &                       &       &             \\[-2.5mm]
\hline\hline
                    &                &                                             &                  &             &      &              &       &                     &                       &       &             \\
\multicolumn{3}{|l|}{\large\it Galactic Center cluster}                            &                  &             &      &              &       &                     &                       &       &             \\[+2mm]
     100a           &                &      GC\,AF\,NW\,NW             \hfill{E81} &      WN7         &             &      & $K$ = 12.6   & PG06  &      17 45 39.306   &      $-$29 00 30.68   & $a$   &      PG06   \\
{\it 101a }         & {\it BSD\,1}   & {\it MPE$-$8.3$-$5.7}           \hfill{E82} & {\it WC9}        &    WC8-9    & PG06 & $K$ = 13.0   & PG06  &      17 45 39.382   &      $-$29 00 33.43   & $a$   & {\it BS95 } \\
{\it 101b }         & {\it KGE\,1}   & {\it AF\,NW}                    \hfill{E74} & {\it WN9-11}     &    WN8      & PG06 & $K$ = 11.7   & PG06  &      17 45 39.458   &      $-$29 00 31.67   & $a$   & {\it KG95 } \\
{\it 101c }         & {\it KGE\,2}   & {\it AF}                        \hfill{E79} & {\it WN9-11}     &    Ofpe/WN9 & PG06 & $K$ = 10.8   & PG06  &      17 45 39.541   &      $-$29 00 35.01   & $a$   & {\it KG95 } \\
{\it 101d }         & {\it KGE\,3}   & {\it GC\,IRS\,6E}                           & {\it WC9}        &             &      & $K$ = ~\,9.55& OE99  &      17 45 39.643   &      $-$29 00 27.33   & $b$   & {\it KG95 } \\
     101da          &      PGM\,1    &                                 \hfill{E60} &      WN7?        &             &      & $K$ = 12.4   & PG06  &      17 45 39.708   &      $-$29 00 29.75   & $a$   &      PG06   \\
     101db          &      PGM\,2    &      GC\,IRS\,34W               \hfill{E56} &      Ofpe/WN9    &             &      & $K$ = 11.4   & PG06  &      17 45 39.731   &      $-$29 00 26.51   & $a$   &      PG06   \\
     101dc          &      MES-WR\,1 &      GC\,IRS\,7SW               \hfill{E66} &      WN8-9       &    WN8      & PG06 & $K$ = 12.0   & PG06  &      17 45 39.739   &      $-$29 00 23.17   & $a$   &      ME05   \\
     101dd          &      PGM\,3    &      GC\,IRS\,34NW              \hfill{E61} &      WN7         &             &      & $K$ = 12.8   & PG06  &      17 45 39.756   &      $-$29 00 25.25   & $a$   &      PG06   \\
     101de          &      MPS\,1    &      GC\,IRS\,13E5                          &      WCLd?       &             &      & $K$ = 11.90  & MP04  &      17 45 39.780   &      $-$29 00 29.65   & $c$   &      MP04   \\
     101df          &      MPS\,2    &      GC\,IRS\,13E3B                         &      WCLd?       &             &      & $K$ = 13.07  & MP04  &      17 45 39.792   &      $-$29 00 29.59   & $c$   &      MP04   \\
     101dg          &      TGM05-1   &      GC\,IRS\,2 , BSD96-45                  &      WCLd?       &             &      & $K$ = 10.34  & BS96  &      17 45 39.792   &      $-$29 00 34.99   & $i$   &      TG05   \\
     101dh          &      MPS\,3    &      GC\,IRS\,13E3A             \hfill{E49} &      WCLd?       &      ?      & PG06 & $K$ = 13.0   & PG06  &      17 45 39.796   &      $-$29 00 29.63   & $c$   &      MP04   \\
     101di          &      MPS\,4    &      GC\,IRS\,13E4              \hfill{E48} &      WC8-9       &    WC9      & PG06 & $K$ = 11.7   & PG06  &      17 45 39.797   &      $-$29 00 29.52   & $a$   &      MP04   \\
{\it 101e} $^{12}$  & {\it KGE\,5}   & {\it GC\,IRS\,13E2} , MPS\,5    \hfill{E51} & {\it WN9-10 +?}  &    WN8      & PG06 & $K$ = 10.8   & PG06  &      17 45 39.801   &      $-$29 00 29.84   & $a$   & {\it KG95 } \\
     101ea          &      EML\,1    &      GC\,IRS\,13E3c                         &      WCLd?       &             &      & $K$ = 12.49  & MP04  &      17 45 39.808   &      $-$29 00 29.48   & EM04  &      EM04   \\
{\it 101f} $^{13}$  & {\it KGE\,4}   & {\it GC\,IRS\,7W} ,  MES-WR2    \hfill{E68} & {\it WN9-10}     &    WC9      & PG06 & $K$ = 13.1   & PG06  &      17 45 39.853   &      $-$29 00 22.11   & $a$   & {\it KG95 } \\
     101fa          &      HET\,1    &      GC\,IRS\,3E                \hfill{E58} &      WC5-6d      &    WC5-6d?  & PE05 & $K$ = 15.0   & PG06  &      17 45 39.868   &      $-$29 00 24.30   & $a$   &      HE04   \\
{\it 101g}          & {\it KGE\,6}   & {\it GC\,IRS\,29N}              \hfill{E31} & {\it WC9}        &             &      & $K$ = 10.0   & PG06  &      17 45 39.918   &      $-$29 00 26.69   & $a$   & {\it KG95 } \\
{\it 101h} $^{14}$  & {\it KGE\,8}   & {\it GC\,IRS\,15SW} , MES-WR3   \hfill{E83} & {\it WN9-11}     &    WN8-WC9  & PG06 & $K$ = 12.0   & PG06  &      17 45 39.920   &      $-$29 00 18.08   & $a$   & {\it KG95 } \\
{\it 101i} $^{\,15}$& {\it KGE\,7}   & {\it GC\,IRS\,29NE1, MPE$-$1.0$-$3.5}  ~E35 & {\it WC9}        &    WC8-9    & PG06 & $K$ = 11.7   & PG06  &      17 45 39.965   &      $-$29 00 26.04   & $a$   & {\it KG95 } \\
{\it 101j}          & {\it KGE\,9}   & {\it GC\,IRS\,16NW}             \hfill{E19} & {\it WN9-11}     &    Ofpe/WN9 & PG06 & $K$ = 10.0   & PG06  &      17 45 40.042   &      $-$29 00 26.89   & $a$   & {\it KG95 } \\
     101ja          &      PGM\,4    &      GC\,IRS\,33E               \hfill{E41} &      Ofpe/WN9    &             &      & $K$ = 10.1   & PG06  &      17 45 40.090   &      $-$29 00 31.22   & $a$   &      PG06   \\
{\it 101k}          & {\it KGE\,10}  & {\it GC\,IRS\,16SW}             \hfill{E23} & {\it WN9-11\,+?} &    Ofpe/WN9 & PG06 &$K$ =\,~~9.61v& GP00  &      17 45 40.120   &      $-$29 00 29.08   & $a$   & {\it KG95 } \\
{\it 101l}          & {\it KGE\,11}  & {\it GC\,IRS\,16C}              \hfill{E20} & {\it WN9-11}     &    Ofpe/WN9 & PG06 & $K$ = \,~9.7 & PG06  &      17 45 40.126   &      $-$29 00 27.62   & $a$   & {\it KG95 } \\
{\it 101m}          & {\it KGE\,12}  & {\it GC\,IRS\,15NE}             \hfill{E88} & {\it WN9-11}     &    WN8-9    & PG06 & $K$ = 11.8   & PG06  &      17 45 40.145   &      $-$29 00 16.42   & $a$   & {\it KG95 } \\
     101ma          &      PGM\,5    &                                 \hfill{E71} &      WC8-9?      &             &      & $K$ = 14.1   & PG06  &      17 45 40.161   &      $-$29 00 21.61   & $a$   &      PG06   \\
{\it 101n}          & {\it KGE\,13}  & {\it GC\,IRS\,16SE1, MPE$+$1.6$-$6.8} \hfill{E32} & {\it WC9}  &    WC8-9    & PG06 & $K$ = 10.9   & PG06  &      17 45 40.181   &      $-$29 00 29.25   & $a$   & {\it KG95 } \\
     101na          &      TGM02-1   &      GC\,IRS\,21 ,   BSD96-81               & WCLd?            &             &      & $K$ = 10.55  & CR01  &      17 45 40.221   &      $-$29 00 30.84   & $i$   &      TG02   \\
     101nb          &      PGM\,6    &      GC\,IRS\,7SE2                          &      WC?         &             &      & $K$ = 13.7   & PG06  &      17 45 40.245   &      $-$20 00 24.23   & $k$   &      PG06   \\
     101nc          &      PGM\,7    &      GC\,IRS\,9W                \hfill{E65} & WN8              &             &      & $K$ = 12.1   & PG06  &      17 45 40.257   &      $-$29 00 33.72   & $a$   &      PG06   \\
     101nd          &      PGM\,8    &      GC\,IRS\,16NE              \hfill{E39} & Ofpe/WN9         &             &      & $K$ = ~\,8.9 & PG06  &      17 45 40.259   &      $-$29 00 27.07   & $a$   &      PG06   \\
{\it 101o}          & {\it KGE\,14}  & {\it GC\,IRS\,16SE2, MPE$+$2.7$-$6.9}  \hfill{E40} & {\it WC9} &    WN5-6    & HE04 & $K$ = 12.0   & PG06  &      17 45 40.264   &      $-$29 00 29.29   & $a$   & {\it KG95 } \\
                    &                &                                             &                  &             &      &              &       &                     &                       &       &             \\
\hline
\end{tabular}
\end{sideways}
\end{table*}
\end{center}

% \clearpage

\begin{center}
\begin{table*}[t!]
\begin{sideways}
\begin{tabular}{|lll|lll|ll|llc|l|}
\hline\hline
\multicolumn{12}{|l|}{}                                                                                                                                                                                               \\[-1.5mm]
\multicolumn{12}{|l|}{{\bf Table 1 (cont'd):} {\large New Galactic Wolf-Rayet stars.
 {\it Data quoted from 7Cat are listed in italics},
 revised and new data are listed in roman font.}}                                                                                                                                                                                     \\
\multicolumn{12}{|l|}{}                                                                                                                                                                                               \\[-1.5mm]
\hline\hline
                    &                &                                            &                   &             &      &              &       &                     &                       &       &             \\[-2.5mm]
  WR                &   WR           & other designation(s)                       &      discovery    &    revised  &      &   $m$        &       & RA(J2000)           & Dec(J2000)            &       & WR          \\
                    &   discovery    &                               \hfill{in}   &      spectral     &    spectral &      &  (mag)       &       &                     &                       &       & discovery   \\
                    &   designation  &                               \hfill{PG06} &      type         &    type     & ref. &              & ref.  &                     &                       & ref.  & ref.        \\
                    &                &                                            &                   &             &      &              &       &                     &                       &       &             \\[-2.5mm]
\hline\hline
                    &                &                                            &                   &             &      &              &       &                     &                       &       &             \\
\multicolumn{3}{|l|}{\large\it Galactic Center cluster (cont'd)}                  &                   &             &      &              &       &                     &                       &       &             \\[+2mm]
     101oa          &      PMM\,1    &      He{\sc i}\,N3             \hfill{E59} &      WR           &    WC9      & PG06 & $K$ = 13.0   & PG06  &      17 45 40.264   &      $-$29 00 24.64   & $a$   & PM01        \\
     101ob          &      PGM\,9    &      GC\,IRS\,9SW \hfill{E76}              &      WC9          &             &      & $K$ = 13.1   & PG06  &      17 45 40.366   &      $-$29 00 36.13   & $a$   & PG06        \\
     101oc          &      PMM\,2    &      GC\,IRS\,7E2\,(ESE)       \hfill{E70} &      WR           &    Ofpe/WN9 & PG06 & $K$ = 12.9   & PG06  &      17 45 40.369   &      $-$29 00 22.76   & $b$   & PM01        \\
     101od          &      TGM05-3   &      GC\,IRS\,5                            &      WCLd?        &             &      &              &       &      17 45 40.4     &      $-$29 00 16      & $d$   & TG05        \\
     101oe          &      MEV\,1    &      GC\,IRS\,1W ,  BSD96-92               &      WCLd?        &             &      & $K$ = \,~8.72& CR01  &      17 45 40.442   &      $-$29 00 27.53   & $i$   & ME04        \\
     101of          &      PGM\,10   &      GC\,IRS\,9SE              \hfill{E80} &      WC9          &             &      & $K$ = 11.7   & PG06  &      17 45 40.471   &      $-$29 00 36.27   & $a$   & PG06        \\
     101og          &      TGM05-4   &      GC\,IRS\,10W ,  BSD96-94              &      WCLd?        &             &      & $K$ = 10.25  & BS96  &      17 45 40.49    &      $-$29 00 22.8    & OE99  & TG05        \\
     101oh          &      PGM\,11   &                                \hfill{E72} &      WC9?         &             &      & $K$ = 13.6   & PG06  &      17 45 40.551   &      $-$29 00 28.60   & $a$   & PG06        \\
     101oi          &      PMM\,3    &      ID\,180 , He{\sc i}\,N1   \hfill{E78} &      WR           &    WC9      & PG06 & $K$ = 13.0   & PG06  &      17 45 40.760   &      $-$29 00 27.79   & $c$   & PM01        \\
                    &                &                                            &                   &             &      &              &       &                     &                       &       &             \\
\hline
                    &                &                                            &                   &             &      &              &       &                     &                       &       &             \\
     101p           &      HBP\,1    &                                            &      WC8-9        &             &      &$K_s$= 11.20  & HB03  &      17 45 42.47    &      $-$28 52 53.3    & HB03  & HB03        \\
                    &                &                                            &                   &             &      &              &       &                     &                       &       &             \\
\hline
                    &                &                                            &                   &             &      &              &       &                     &                       &       &             \\
\multicolumn{3}{|l|}{\large\it Arches cluster}                                    &                   &             &      &              &       &                     &                       &       &             \\[+2mm]
{\it 102a}          & {\it CSE\,1}   & {\it ``\,near G\,0.10+0.20"        }       & {\it WN8      }   &             &      & $K'$= 10.22  & CS99  &      17 45 48.560   &      $-$28 50 06.08   & $f$   & {\it CS99 } \\
     102aa          &      NWS\,1    &      C13 ,   AR6 ,~   B34 ,  F2            &      WN9          &    WN9+OB?  & LG01 & $K'$= 10.7~  & CE96  &      17 45 49.76    &      $-$28 49 26.0    & LJ05  & NW95        \\
     102ab          &      BSP\,30   &     ~~~~~~~~~~~~~~~~~~B30 ,  F10           &      WN7          &             &      & $K'$= 11.46  & FNO2  &      17 45 50.08    &      $-$28 49 26.2    & $g$   & BS01        \\
     102ac          &      BSP\,29   &     ~~~~~~~~~~~~~~~~~~B29 ,  F17           &      WN7          &             &      & $K'$= 12.15  & FN02  &      17 45 50.15    &      $-$28 49 26.9    & $g$   & BS01        \\
     102ad          &      NWS\,4    &      C9 ,~   AR3 ,~\, B28 ,  F1            &      WN9          &    WN9+OB?  & LJ05 & $K'$= 10.2~  & CE96  &      17 45 50.20    &      $-$28 49 22.3    & LJ05  & NW95        \\
     102ae          &      NWS\,5    &     C1 ,~~~~~~~~~~~\, B26 ,  F9            &      WN9          &             &      & $K'$= 10.6~  & CE96  &      17 45 50.31    &      $-$28 49 11.5    & $g$   & NW95        \\
     102af          &      NWS\,6    &      C3 ,~   AR16 ,   B25 ,  F12           &      WN9          &             &      & $K'$= 10.6~  & CE96  &      17 45 50.31    &      $-$28 49 17.0    & $g$   & NW95        \\
     102ag          &      NWS\,7    &      C6 ,~   AR2 ,~   B24 ,  F8            &      WN9          &             &      & $K'$= 10.76  & FN02  &      17 45 50.39    &      $-$28 49 21.3    & LJ05  & NW95        \\
     102ah          &      NWS\,8    &      C8 ,~   AR1 ,~   B23 ,  F6            &      WN9          &    WN9+OB?  & LJ05 & $K'$= 10.1~  & CE96  &      17 45 50.42    &      $-$28 49 22.3    & LJ05  & NW95        \\
     102ai          &      NWS\,9    &       ~~~~~~~AR8 ,~\, B22 ,  F5            &      WN9          &    WN9+OB?  & LJ05 & $K'$= 10.86  & FN02  &      17 45 50.45    &      $-$28 49 31.9    & LJ05  & NW95        \\
     102aj          &      NWS\,10   &      C5 ,~   AR4 ,~   B21 ,  F7            &      WN9          &    WN9+OB?  & LJ05 & $K'$= \,~9.7~& CE96  &      17 45 50.47    &      $-$28 49 19.5    & LJ05  & NW95        \\
     102ak          &      BSP\,19   &    ~~~~~~~~~~~~~~~~~\,B19 ,  F16           &      WN6-7        &             &      & $K'$= 11.40  & FN02  &      17 45 50.55    &      $-$28 49 20.5    & $g$   & BS01        \\
     102al          &      NWS\,11   &      C2 ,~   AR5 ,~   B17 ,  F4            &      WN9          &    WN8      & La03 & $K'$= 10.2~  & CE96  &      17 45 50.57    &      $-$28 49 17.5    & LJ05  & NW95        \\
{\it 102b}          & {\it CSE\,2}   & {\it ``\,near Sgr\,A\,East region A" }     & {\it WN6      }   &             &      & $K'$= 10.97  & CS99  &      17 45 50.626   &      $-$28 59 19.61   & $f$   & {\it CS99 } \\
     102ba          &      CEC\,7    &     ~~~~~~~~~~~~~~~~~~B12 ,  F14           &      WN7          &             &      & $K'$= 11.22  & FN02  &      17 45 50.69    &      $-$28 49 22.5    & $g$   & CE96        \\
     102bb          &      NWS\,14   &      C11 ,   AR7 ,~   B3  ,~ F3            &      WN9          &    WN9/Ofpe & La03 & $K'$= 10.3~  & CE96  &      17 45 50.83    &      $-$28 49 26.4    & LJ05  & NW95        \\
     102bc          &      CEC\,10   &     ~~~~~~~~~~~~~~~~~~B1                   &      WN7          &             &      & $K'$= 11.3   & CE96  &      17 45 51.46    &      $-$28 49 26.0    & $g$   & CE96        \\
                    &                &                                            &                   &             &      &              &       &                     &                       &       &             \\
\hline
\end{tabular}
\end{sideways}
\end{table*}
\end{center}

% \clearpage

\begin{center}
\begin{table*}[t!]
\begin{sideways}
\setlength{\tabcolsep}{1.7mm}
\begin{tabular}{|lll|lll|ll|llc|l|}
\hline\hline
\multicolumn{12}{|l|}{}                                                                                                                                                                                               \\[-1.5mm]
\multicolumn{12}{|l|}{{\bf Table 1 (cont'd):} {\large New Galactic Wolf-Rayet stars.
 {\it Data quoted from 7Cat are listed in italics},
      revised and new data are listed in roman font.}}                                                                                                                                                                \\
\multicolumn{12}{|l|}{}                                                                                                                                                                                               \\[-1.5mm]
\hline\hline
                    &                &                                        &                       &             &      &             &       &                     &                       &       &              \\[-2.5mm]
  WR                &   WR           & other                                  &      discovery        &    revised  &      &   $m$       &       & RA(J2000)           & Dec(J2000)            &       & WR           \\
                    &   discovery    & designation(s)                         &      spectral         &    spectral &      &  (mag)      &       &                     &                       &       & discovery    \\
                    &   designation  &                                        &      type             &    type     & ref. &             & ref.  &                     &                       & ref.  & ref.         \\
                    &                &                                        &                       &             &      &             &       &                     &                       &       &              \\[-2.5mm]
\hline\hline
                    &                &                                        &                       &             &      &             &       &                     &                       &       &              \\
     102bd  $^{16}$ &      HBP\,2    &                                        &      WC8-9            &             &      &$K_s$= 11.49 & HB03  &      17 45 57.78    &      $-$28 54 46.1    & HB03  & HB03         \\
                    &                &                                        &                       &             &      &             &       &                     &                       &       &              \\
\hline
                    &                &                                        &                       &             &      &             &       &                     &                       &       &              \\
\multicolumn{3}{|l|}{\large\it Quintuplet cluster}                            &                       &             &      &             &       &                     &                       &       &              \\[+2mm]
{\it 102c}          & {\it FMM96-1 } & {\it ~~~~~~~~~~~~\,qF353E}             & {\it WN6          }   &             &      & $K$ =11.53  & FM99  & {\it 17 46 11.2   } & {\it $-$28 49 05.6  } & 7Cat  & {\it FM95 }  \\
     102ca          &      HBP\,3    &                                        &      WC8-9            &             &      & $K_s$=10.40 & HB03  &      17 46 13.04    &      $-$28 49 25.4    & HB03  & HB03         \\
{\it 102d}          & {\it FMM95-1 } & {\it ~~~~~~~~~~~~\,qF320} ,~~~~~~~~QR8 & {\it WN9          }   &             &      & $K$ =10.50  & FM99  &      17 46 14.067   &      $-$28 49 17.28   & $h$   & {\it FM95 }  \\
     102da          &      FMM-d1    &      GCS\,3-4 ,    qF243 ,   Q1        &      WCLd?            &             &      & $K$ =~7.61  & GM99  &      17 46 14.151   &      $-28$ 49 37.42   & $h$   & FM96         \\
     102db          &      FMM-d2    &      GCS\,3-3 ,    qF258 ,   Q9        &      WCLd?            &             &      & $K$ =~8.98  & GM99  &      17 46 14.336   &      $-28$ 49 32.17   & $h$   & FM96         \\
     102dc          &      FMM-d3    &      GCS\,3-2 ,    qF231 ,   Q2 ,  QR7 &      WCLd?            &  WC7-8d+OB  & TM06 & $K$ =~6.28v & GM99  &      17 46 14.721   &      $-28$ 49 41.46   & $h$   & FM96         \\
     102dd          &      FMM-d4    &      GCS\,3-1 ,    qF251 ,   Q4        &      WCLd?            &             &      & $K$ =~7.66  & GM99  &      17 46 14.810   &      $-28$ 49 35.02   & $h$   & FM96         \\
{\it 102e}          & {\it FMM96-2 } & {\it ~~~~~~~~~~~~\,qF151}              & {\it WC8          }   &             &      & $K$ =10.44  & FM99  &      17 46 14.827   &      $-$28 50 01.17   & $h$   & {\it FM96 }  \\
     102ea          &      FMM96-7   &     ~~~~~~~~~~~~~\,qF241 ,   Q10,  QR5 &      WN9/Ofpe         &             &      & $K$ =~8.83  & GM99  &      17 46 15.129   &      $-$28 49 37.82   & $j$   & FM96         \\
{\it 102f}          & {\it FMM96-3 } & {\it ~~~~~~~~~~~~\,qF235N}             & {\it WC$<$8\,+\,? }   &             &      &             &       &      17 46 15.168   &      $-$28 49.40.25   & $h$   & {\it FM96 }  \\
{\it 102g}          & {\it FMM99-1 } & {\it ~~~~~~~~~~~~\,qF235S}             & {\it WC$<$8       }   &             &      &             &       &      17 46 15.182   &      $-$28 49 42.40   & $h$   & {\it FM99 }  \\
{\it 102h}          & {\it FMM95-2 } & {\it ~~~~~~~~~~~~\,qF76}               & {\it WC9          }   &             &      & $K$ =11.44  & FM99  &      17 46 15.572   &      $-$28 50 18.89   & $h$   & {\it FM95 }  \\
     102ha          &      FMM-d5    &      GCS\,4 ,~~    qF211 , \,Q3        &      WCLd?            &    WCLd+OB  & TM06 & $K$ =~6.91v & GM99  &      17 46 15.884   &      $-28$ 49 46.27   & $h$   & FM96         \\
     102hb          &      FMM96-8   &     ~~~~~~~~~~~~~\,qF240 ,   Q8        &      WN9/Ofpe         &             &      & $K$ =~9.01  & GM99  &      17 46 15.954   &      $-$28 49 38.60   & $j$   & FM96         \\
{\it 102i}          & {\it FMM96-4 } & {\it ~~~~~~~~~~~~\,qF256}              & {\it WN9\,+\,?    }   &             &      & $K$ =11.38  & FM99  &      17 46 16.560   &      $-$28 49 32.53   & $h$   & {\it FM96 }  \\
{\it 102j} $^{17}$  & {\it FMM96-6 } & {\it ~~~~~~~~~~~~\,qF309}              & {\it WC$<$8       }   &             &      & $K$ =11.52  & FM99  &      17 46 17.522   &      $-$28 49 19.41   & $h$   & {\it FM96 }  \\
{\it 102k} $^{18}$  & {\it FMM96-5 } & {\it ~~~~~~~~~~~~\,qF274}              & {\it WN9          }   &             &      & $K$ =11.41  & FM99  &      17 46 17.548   &      $-$28 49 29.52   & $h$   & {\it FM96 }  \\
                    &                &                                        &                       &             &      &             &       &                     &                       &       &              \\
\hline
                    &                &                                        &                       &             &      &             &       &                     &                       &       &              \\
     102ka          &      HBP\,4    &                                        &      WN10             &             &      & $K_s$=~8.84 & HB03  &      17 46 18.12    &      $-$29 01 36.5    & HB03  & HB03         \\
                    &                &                                        &                       &             &      &             &       &                     &                       &       &              \\
\hline
                    &                &                                        &                       &             &      &             &       &                     &                       &       &              \\
\multicolumn{3}{|l|}{\large\it ``\,SGR\,1806$-$20" cluster}                   &                       &             &      &             &       &                     &                       &       &              \\[+2mm]
     111a           &      FNG\,1    &      FNG-1                             &      WC8              &             &      & $K$ =11.76  & FN05  &      18 08 38.32    &      $-$20 24 33.5    & FN05  & FN05         \\
     111b           &      EGH\,1    &      FNG-B                             &      WC9d             &             &      & $K$ =10.50  & FN05  &      18 08 39.24    &      $-$20 24 42.50   & FN05  & EG01, EML04  \\
     111c           &      FNG\,2    &      FNG-2                             &      WN6              &             &      & $K$ =12.16  & FN05  &      18 08 39.42    &      $-$20 24 42.57   & FN05  & FN05         \\
     111d           &      FNG\,3    &      FNG-3                             &      WN7?             &             &      & $K$ =12.87  & FN05  &      18 08 39.50    &      $-$20 24 35.88   & FN05  & FN05         \\
                    &                &                                        &                       &             &      &             &       &                     &                       &       &              \\
\hline
                    &                &                                        &                       &             &      &             &       &                     &                       &       &              \\
     142a           &      PCG\,1    &      NGC\,6910-MS\,21                  &      WC8              &             &      & $K_s$=~7.09 & PC02  &      20 24 06.2     &      $+$41 25 33      & PC02  & PC02         \\[-2mm]
                    &                &                                        &                       &             &      &             &       &                     &                       &       &              \\
     159            &      BCC\,1    &      BD+62$^\circ$2296B                &      WN4              &             &      & $V_T$=11.20 & Ne03  &      23 47 20.4     &      $+$63 13 14      & Ne03  & BC94, Ne03   \\[-2mm]
                    &                &                                        &                       &             &      &             &       &                     &                       &       &              \\
\hline\hline
\end{tabular}
\end{sideways}
\end{table*}
\end{center}

\clearpage

\noindent
{\bf Notes to Table 1}                \\

\footnotesize

\noindent
{\it Revised WR numbers of stars in 7Cat:}

\noindent
$^1$   : WR\,77c: formerly WR\,77b in NC03.   \\
$^2$   : WR\,77e: formerly WR\,77a in NC03.   \\
$^3$   : WR\,77g: formerly WR\,77c in NC03.   \\
$^4$   : WR\,77j: formerly WR\,77e in NC03.   \\
$^5$   : WR\,77l: formerly WR\,77d in NC03.   \\
$^6$   : WR\,77m: formerly WR\,77f in NC03.   \\
$^7$   : WR\,77n: formerly WR\,77g in NC03.   \\
$^{8}$ : WR\,77o: formerly WR\,77h in NC03.   \\
$^{9}$ : WR\,77p: formerly WR\,77i in NC03.   \\
$^{10}$: WR\,77r: formerly WR\,77j in NC03.   \\
$^{11}$: WR\,77sc: formerly WR\,77k in NC03.  \\
$^{12}$: {\it WR\,101e}: formerly {\it WR\,101f} in 7Cat. \\
\hspace*{4mm} Erratum: for GC\,IRS\,13E1 in 7Cat, read GC\,IRS\,13E2. \\
$^{13}$: {\it WR\,101f}: formerly {\it WR\,101e} in 7Cat.  \\
$^{14}$: {\it WR\,101h}: formerly {\it WR\,101i} in 7Cat.  \\
$^{15}$: {\it WR\,101i}: formerly {\it WR\,101h} in 7Cat.  \\
$^{16}$: {\it WR\,102bd}: formerly {\it WR\,101q} in HB03.  \\
$^{17}$: {\it WR\,102j}: formerly {\it WR\,102k} in 7Cat.  \\
$^{18}$: {\it WR\,102k}: formerly {\it WR\,101j} in 7Cat.  \\

\noindent
{\it Magnitudes:      }

\noindent
For each object the most recently published magnitude has been quoted,
unless the new observation only confirms the earlier observation. 
CS99 used $K'$($\lambda$$_c$ = 2.11\,$\mu$m).   
FN02 used $m_{\rm F205W}$.                      
HB03 used $K_s$ (narrow continuum filter $\lambda$$_c$ = 2.248\,$\mu$m) from 2MASS. \\

\noindent
{\it Coordinates: }

\noindent
Coordinates from reference in last column, unless indicated otherwise
(p.c. = private communication):                                         \\
$a$: coordinates from F. Martins, 11 August 2005, p.c.;                 \\
\hspace*{3mm} also PG05.                                                \\
$b$: revised coordinates from T. Paumard, October 2004,                 \\
\hspace*{3mm}  p.c.                                                     \\
$c$: coordinates from T. Paumard, August 2004, p.c.                     \\
$d$: coordinates from CDS-Simbad.                                       \\
$e$: coordinates from J. Moultaka, August 2005, p.c.                    \\
$f$: coordinates from A.S. Cotera, July 2005, p.c.                      \\
$g$: coordinates from R.D. Blum, August 2004, p.c.                      \\
$h$: coordinates from D.F. Figer, August 2004, p.c.                     \\
$i$: coordinates from F. Martins, 30 August 2005, p.c.;                 \\
\hspace*{3mm} also PG05.                                                \\
$j$: coordinates from D.F. Figer, April 2006, p.c.                      \\
$k$: coordinates from F. Martins, May 2006, p.c.                        \\

\noindent
{\it Reference abbreviations:}

\noindent
AR : Lang \ea \cite{lan2001}; Lang \cite{lan2003},
     Lang \ea \cite{lan2005}                                        \\
BC94 = BCC          : Bartaya \ea \cite{bar1994}                    \\
BS01 = B = BSP      : Blum \ea \cite{blu2001}                       \\
BS95 = BSD          : Blum \ea \cite{blu1995}                       \\
BS96 = BSD96        : Blum \ea \cite{blu1996}                       \\
CE96 = C = CEC      : Cotera \ea \cite{cot1996}                     \\
CH06                : Crowther \ea \cite{cro2006}                   \\
CN02                : Clark \& Negueruela \cite{cla2002}            \\
CP05 = CPG          : Cohen \ea \cite{coh2005}                      \\
CR01                : Cl\'enet \ea \cite{cle2001}                   \\
CS99                : Cotera \ea \cite{cot1999}                     \\
DB04 = DBU          : Drew \ea \cite{dre2004}                       \\
E                   : running number in Paumard \ea {\cite{pau2006}, Table\,2   \\
EG01                : Eikenberry \ea \cite{eik2001}                 \\
EM04                : Eckart \ea \cite{eck2004}                     \\
EML  = EML04        : Eikenberry \ea \cite{eik2004}                 \\
F                   : Figer \ea \cite{fig2002}                      \\
FM95                : Figer \ea \cite{fig1995}                      \\
FM96                : Figer \ea \cite{fig1996}                      \\
FM99a = FMM = FMM99 : Figer \ea \cite{fig1999a}                     \\
FM99b = FMG99       : Figer \ea \cite{fig1999b}                     \\
FN05 = FNG          : Figer \ea \cite{fig2005}                      \\
GCS                 : Nagata \ea \cite{nag1995}                     \\
GD06 = GDTB         : Groh \ea \cite{gro2006}                       \\
GM99                : Glass \ea \cite{gla1999}                      \\
GP00 = GPE          : Genzel \ea \cite{gen2000}                     \\
HB03 = HBP          : Homeier \ea \cite{hom2003}                    \\
HB05 = HBD          : Hopewell \ea \cite{hop2005}                   \\
HE04 = HET          : Horrobin \ea \cite{hor2004}                   \\
KG95 = KGE          : Krabbe \ea \cite{kra1995}                     \\
La03                : Lang \cite{lan2003}                           \\
LG01                : Lang \ea \cite{lan2001}                       \\
LJ05                : Lang \ea \cite{lan2005}                       \\
ME04 = MEV          : Moultaka \ea \cite{mou2004}                   \\
ME05 = MES          : Moultaka \ea \cite{mou2005}                   \\
MP04 = MPS          : Maillard \ea \cite{mai2004}                   \\
NC03                : Negueruela \& Clark \cite{ncl2003}            \\
NC05 = NC           : Negueruela \& Clark \cite{ncl2005}            \\
Ne03                : Negueruela \cite{neg2003}                     \\
Ne05                : Negueruela, priv. comm.: VLT-FORS spectroscopy \\
NW95 = NWS          : Nagata \ea \cite{nag1995}                     \\
OE99                : Ott \ea \cite{ott1999}                        \\
Pa04                : Paumard 2004, private communication           \\
PC02 = PCG          : Pasquali \ea \cite{pas2002}                   \\
PE05                : Pott \ea \cite{pot2005}                       \\
PG04                : Paumard \ea \cite{pau2004}                    \\
PG05                : Paumard \ea \cite{pau2005}                    \\
PG06 = PGM          : Paumard \ea \cite{pau2006}                    \\
PM01 = PMM          : Paumard \ea \cite{pau2001}                    \\
PM03                : Paumard \ea \cite{pau2003}                    \\
Q = GMM = GM90      : Glass \ea \cite{gla1990} (see also Moneti \ea \\
                      \hspace*{6mm} \cite{mon2001})                 \\
QR                  : Lang \ea \cite{lan1999}; Lang \cite{lan2003};
                      Lang \ea \cite{lan2005}                       \\
TG02 = TGM02        : Tanner \ea \cite{tan2002}                     \\
TG05 = TGM05        : Tanner \ea \cite{tan2005}                     \\
TM06                : Tuthill \ea \cite{tut2006}                    \\

\section{Notes on individual stars}

\smallskip \noindent
{\large\it Westerlund\,1 :}

\smallskip \noindent
{\bf 77b} = NC-N                : X-ray detection by {\sl Chandra}
                                  (Skinner \ea \cite{ski2006}).

\smallskip \noindent
{\bf 77g} = NC-K                : X-ray detection by {\sl Chandra}
                                  (Skinner \ea \cite{ski2006}).

\smallskip \noindent
{\bf 77j} = NC-G                : X-ray detection by {\sl Chandra}
                                  (Skinner \ea \cite{ski2006}).

\smallskip \noindent
{\bf 77k} = NC-L = Wd1-44       : X-ray detection by {\sl Chandra}
                                  (Skinner \ea \cite{ski2006}).

\smallskip \noindent
{\bf 77n} = NC-F = Wd1-239      : X-ray detection by {\sl Chandra}
                                  (Skinner \ea \cite{ski2006}).

\smallskip \noindent
{\bf 77o} = NC-B                : X-ray detection by {\sl Chandra}
                                  (Skinner \ea \cite{ski2006}).
                                  Relatively high $L_{\rm x}$,
                                  possibly colliding wind binary.

\smallskip \noindent
{\bf 77p} = NC-E = Wd1-241      : X-ray detection by {\sl Chandra}
                                  (Skinner \ea \cite{ski2006}).

\smallskip \noindent
{\bf 77q} = NC-R = WD1-14c      : X-ray detection by {\sl Chandra}
                                  (Skinner \ea \cite{ski2006}).

\smallskip \noindent
{\bf 77r} = NC-D                : X-ray detection by {\sl Chandra}
                                  (Skinner \ea \cite{ski2006}).

\smallskip \noindent
{\bf 77sa} = NC-W = GDTB\,3     : X-ray detection by {\sl Chandra}
                                  (Skinner \ea \cite{ski2006}).

\smallskip \noindent
{\bf 77sb} = NC-O               : X-ray detection by {\sl Chandra}
                                  (Skinner \ea \cite{ski2006}).

\smallskip \noindent
{\bf 77sc} = NC-A               : X-ray detection by {\sl Chandra}
                                  (Skinner \ea \cite{ski2006}).
                                  Relatively high $L_{\rm x}$,
                                  possibly colliding-wind binary.

\bigskip \noindent
{\large\it Galactic Center cluster :}

\smallskip \noindent
{\bf WR\,101b} = AF-NW          : tentative association with X-ray source
                                  CXOGC/J174539.4$-$2900310 (Baganoff \ea
                                  \cite{bag2003}).

\smallskip \noindent
{\bf WR\,101db} = GC\,IRS\,34W  : irregular variable ($\Delta$$K$\,=\,1.5 mag),
                                  possibly indicative for LBV phase
                                  (Trippe \ea \cite{tri2006}).

\smallskip \noindent
{\bf WR\,101fa} = GC\,IRS\,3    : an ESO VLTI-{\sc midi} observation by
                                  Pott \ea (\cite{pot2005}) shows a $N$-band
                                  (8-12 $\mu$m) size of $\le$\,40\,mas, i.e.,
                                  $\le$\,300\,AU, compatible with the typical
                                  dust envelope size of WCd stars (Williams
                                  \ea \cite{wil1987}). However, Pott \ea
                                  argue that the WC5-6 spectrum may be
                                  associated with a faint star $\sim$\,120\,mas
                                  east of IRS\,3. See also Viehmann \ea
                                  (\cite{vie2006}).

\smallskip \noindent
{\bf WR\,101k} = GC\,IRS\,16SW  : periodic IR variable, $K$-band light curve,
                                  $P$\,=\,9.725\,d, $M$\,$\simeq$\,100\,M$_\odot$
                                  (Ott \ea \cite{ott1999}; De Poy \ea
                                  \cite{dep2004}).

\smallskip \noindent
{\bf WR\,101nd} = GC\,IRS\,16NE : RV variable, may be SB (Tanner \ea
                                  \cite{tan2006}).

\bigskip \noindent
{\large\it Arches cluster :}

\smallskip \noindent
{\bf WR\,102aa} = NWS\,1  = AR6 : non-thermal radio source (Lang \ea
                                  \cite{lan2001}). 
                                  X-ray detection (Wang \ea \cite{wan2006}).
                                  Maybe WN8+OB colliding wind
                                  binary.

\smallskip \noindent
{\bf WR\,102ad} = NWS\,4  = AR3 : moderately variable (29\%) radio source,
                                  possibly indicative of a colliding wind
                                  binary (Lang \ea \cite{lan2005}).

\smallskip \noindent
{\bf WR\,102ae} = NWS\,5        : source A2 in X-ray detection by 
                                  Law \& Yusef-Zadeh (\cite{law2004}) and
                                  Wang \ea (\cite{wan2006}).

\smallskip \noindent
{\bf WR\,102ah} = NWS\,8  = AR1 : source A1S in X-ray detection by Law \&
                                  Yusef-Zadeh (\cite{law2004})
                                  and Wang \ea (\cite{wan2006}).          
                                  Non-thermal and
                                  moderately variable (12\%) radio source,
                                  possibly indicative of a colliding
                                  wind binary (Lang \ea \cite{lan2005}).

\smallskip \noindent
{\bf WR\,102ai} = NWS\,9  = AR8 : moderately variable (25\%) radio source,
                                  possibly indicative of a colliding wind
                                  binary (Lang \ea \cite{lan2005}).

\smallskip \noindent
{\bf WR\,102aj} = NWS\,10 = AR4 : source A1N in X-ray detection by Law \& Yusef-Zadeh
                                  (\cite{law2004}) and Wang \ea (\cite{wan2006}). 
                                  Moderately variable (30\%)
                                  radio source, possibly indicative of a
                                  colliding wind binary (Lang \ea
                                  \cite{lan2005}).

\smallskip \noindent
{\bf WR\,102b} = Sgr\,A-A       : X-ray detection (Muno \ea \cite{mun2006}).

\bigskip \noindent
{\large\it Quintuplet cluster :}

\smallskip \noindent
{\bf WR\,102dc} = Q2 = GCS3-2 = qF231 = QR7 : 
                                  variable at $K$ (Glass \ea \cite{gla1999},
                                  \cite{gla2001}), indicative of WCLd+OB
                                  colliding wind binary.
                                  Detection in X-rays (Law \& Yusef-Zadeh
                                  \cite{law2004}; Wang \ea (\cite{wan2006}).
                                  IR pinwheel discovered (Tuthill \ea
                                  \cite{tut2006}), proving a WCL+OB colliding
                                  wind binary.

\smallskip \noindent
{\bf WR\,102ha} = Q3 = GCS4 = qF211 : 
                                  variable at $K$ (Glass \ea \cite{gla1999},
                                  \cite{gla2001}), indicative of WCLd+OB
                                  colliding wind binary.
                                  X-ray detection (Wang \ea (\cite{wan2006}).  
                                  Rotating IR pinwheel discovered (Tuthill \ea
                                  \cite{tut2006}), proving WC7-8+OB colliding
                                  wind binary with $P$\,=\,850\,$\pm$\,110\,d.
\\

\normalsize

\section{Conclusion}

The past five years have seen the number of known Galactic WR stars
increase by $\sim$\,30\% to close to 300 objects.  It is to be expected
that, with the advance of observing capabilities, that number will continue
to increase.  Whether the expected number of $\sim$\,1600 WR stars in our
observable quadrant of the Galaxy (van der Hucht \cite{huc2001}) will be
reached remains to be seen.  

Discovering and monitoring WR star in the Galaxy and in the Local Group is 
important for the study of Galactic structure and chemical evolution, and it 
is likely that some WR stars are Type Ib/c supernova progenitors and/or GRB 
progenitors. Identifying even one such object before it explodes could 
contribute greatly to our understanding of these energetic phenomena.

\begin{acknowledgements}
The author is much indebted to Drs.~Bob Blum, Angela Cotera, Paul Crowther,
Don Figer, Ella Hopewell, Jessica LaVine, Fabrice Martins, Jihane Moultaka,
and Thibaut Paumard for providing data on new WR stars in advance of
publication, for re-determining coordinates of 7Cat WR stars in crowded
regions, and for helpful comments and suggestions. Constructive comments 
and suggestions from the referee are highly appreciated. 
\end{acknowledgements}

\end{document}